# Structure, ferromagnetic, dielectric and electronic features of the $LaBiFe_2O_6$ material

J.A. Cuervo Farfán[1], D.M. Aljure García[1], R. Cardona[1,2], J. Arbey Rodríguez[2], D. A. Landínez Téllez[1], J. Roa-Rojas[1]

*1. Grupo de Física de Nuevos Materiales, Departamento de Física, Universidad Nacional de Colombia, AA 5997, Bogotá DC, Colombia*

*2. Grupo de Estudios de Materiales GEMA, Departamento de Física, Universidad Nacional de Colombia, Bogotá DC, Colombia*

57-1-3165000 Ext. 13032

57-1-3165135

jroar@unal.edu.co

**Abstract**

In this paper the synthesis and study of the structural, morphological, electrical, magnetic and electronic properties of the $LaBiFe_2O_6$ novel material are reported. The material was produced using the standard ceramic method. The Rietveld analysis of experimental data of x-ray diffraction showed that it synthesizes in an orthorhombic perovskite structure (*Pnma* space group, # 62). Two types of grain, micro and submicrometric, with the $LaBiFe_2O_6$ stoichiometry were identified by scanning electron microscopy and X-ray dispersive spectroscopy. Results of electrical polarization and dielectric constant reveal the occurrence of hysteretic loops of polarization with evidences of dielectric loss. At room temperature the material is ferromagnetic and exhibits an anomaly at $T=258\ K$, which is attributed to anisotropy effects. Results of diffuse reflectance suggest a semiconductor feature with energy gap $E_g=2,17\ eV$, which is in agreement with calculations of band structure and density of states for one spin orientation, while for the other spin configuration from the calculations a conductor behaviour is expected.

## 1 Introduction

The development of new perovskite-like materials has led to a frantic search of useful physical properties in the industry of solar cells [1], catalytic sensors [2] and spintronics [3], among others. In particular, the ferromagnetic perovskites are extensively examined because its potential prospects of applications in devices to store information on magnetic media, such as the well-known magnetic random access memories [4]. On the other hand, ferroelectric perovskites have been being studied for several decades because of the formation of permanent electric dipoles which by its mutual interaction, give rise to a spontaneous polarization associated with structural deformations related to the positions of the cations and anions within the crystallographic cell unit [5]. These effects are closely linked with the piezoelectric response in these materials, facilitating its implementation on devices transducers from mechanical stress into electrical energy and vice versa [6]. Additionally, due to the high value of their dielectric constant, these materials enable the miniaturization of technological devices in which high polarizations are required under the application of small electric fields [7].

When materials exhibit simultaneously ferromagnetic (or antiferromagnetic) and ferroelectric responses are known as multiferroics [8]. If also these materials evidence coupling between the ferro-type order parameters, in the property known as magnetoelectricity, the potential applicability in spintronic devices for control of magnetic domains and electric dipole moments through the manipulation of small electric fields with low energy dissipation becomes even more plausible [9]. However, it is not easy to find materials in which these properties coexist at room temperature. Perovskites are ceramic oxide materials, characterized by the ideal formula $ABX_3$, where A is an alkali cation, rare earth or transition element, whose ionic radius is larger than that of the cation B, which can be a transition metal, and X is usually Oxygen [10].

The $LaFeO_3$ is a material that has been extensively studied and belongs to this family of ceramic oxides. This crystallizes in an orthorhombic structure at room temperature and therefore is known as orthoferrite [11]. Its most interesting features are: a) its antiferromagnetic response, due to the collinear arrangement of two pseudocubic subnets, consisting of octahedral $FeO_6$ in a magnetic symmetry type G, but additionally evidences weak ferromagnetic response as a result of canted spin order in some cations $Fe^{3+}$ [12]; and b) its electrical polarizability, with a high dielectric constant at low frequencies, caused by oxygen vacancies and the high capacitance in parallel with a high resistance, formed by contact between grains of the polycrystalline system [13]. Moreover, the $BiFeO_3$ perovskite is one of the most studied multiferroics materials [14] because it exhibits antiferromagnetism with Néel temperature $T_N > 350\ °C$ and ferroelectricity with Curie temperature $T_C$ above $800\ °C$ [15]. The structure of this perovskite is rhombohedral at room temperature, which remains until temperatures higher than *600 °C* [16].

Due to the multiferroic nature of the $LaFeO_3$ and $BiFeO_3$ simple perovskites at room temperature, even with structures belonging to different space groups (*Pbnm* and *R3c*, respectively [17,18]) it is interesting to study the $LaFeO_3$ with inclusions of $Bi^{3+}$ in the $La^{3+}$ sites. It was recently reported a study of structural, magnetic and electrical properties of $La_{1-x}Bi_xFeO_3$ ($0 \leq x \leq 0{,}2$), with orthorhombic structure, *Pbnm* space group for all concentrations x, with a slight linear variation of the tolerance factor as a function of concentration [19], but without deepen in possible structural distortions that, most likely, will be associated with detectable changes in the electrical and magnetic properties of the material.

Complex perovskites of the type $AA'B_2O_6$ suggest crystallizing the material in a superstructure of subcells in which cations A and A' must be alternated to form a cell consisting of 8 subcells, so that cations B, will appear octahedrally coordinated with the oxygens. For a perovskite of these characteristics, the tolerance factor is given by the expression

$$\tau = \frac{\frac{r_A + r_{A'}}{2} + r_O}{\sqrt{2}\left(r_B + r_O\right)}, \quad (1)$$

where $r_A$, $r_{A'}$, $r_B$ and $r_O$ represent the ionic radii of A, A', B and O ions. According to the group theoretical analysis of octahedral tilting in ferroelectric perovskites [20], the space groups *Pbnm* of the $LaFeO_3$ and *R3c* of the $BiFeO_3$ come from transitions $Pm\bar{3}m \rightarrow P4/mbm$ (or *Imma*) $\rightarrow Pbnm$ and $Pm\bar{3}m \rightarrow R\bar{3}c$ (or *R3m*) $\rightarrow$ *R3c*, respectively, involving distortions in the crystal structure, caused by displacement of the respective ferroelectric cations $Fe^{3+}$ [21].

The aim of this paper is to analyze extensively the effects of including 50% of the cation $Bi^{3+}$ on the site of $La^{3+}$ on the crystal structure of $LaFeO_3$ in order to form the complex perovskite $LaBiFe_2O_6$, studying possible structural distortions as well as its influence on the magnetic and electrical responses in this material.

# 2 Experimental

The synthesis of the samples was carried out through the implementation of the standard solid reaction technique. First, powders of $La(OH)_3$, $Bi_2O_3$ and $Fe_2O_3$, high purity precursor oxides, were subjected to a temperature of *180 °C* in order to remove moisture. Next, considering the production of samples of *1,0 g* mass, oxides were weighed in stoichiometric proportions. The powders were mixed and milled in an agate mortar for three hours and were subjected to a first heat treatment at a temperature of *550 °C* for *220 h*. Due the low melting temperature of bismuth, the heat treatment was carried out at temperatures below its melting temperature for a long time. Subsequently, the mixture was re-milled for half hour and pressed as a cylindrical pellet of *0,9 cm* diameter to be sintered at *600 °C* for *120 h*. This procedure was repeated to sinter the sample at *715 °C, 740 °C, 770 °C, 800 °C* and *835 °C* for *48 h* each treatment. In order to analyze the crystalline structure of the material, X-ray diffraction (XRD) experiments were carried out in all stages of synthesis, for which a PANalytical X'pert PRO MPD equipment with Cu anode and wavelength 1.54056 Å was used. The experimental data were refined by using the GSAS+EXPGUI code [22], which allows the direct taking of the most important structural parameters and the respective space group. The surface morphology was examined by means a VEGA 3 TESCAN scanning electron microscope (SEM). Semiquantitative composition of the material was studied using the electron dispersive X-ray spectroscopy (EDS) probe coupled to the electron microscope. The magnetization experiments were performed on Quantum Design VersaLab™ MPMS equipment with VSM option. The temperature dependence of the susceptibility was recorded as a function of temperature in magnetic fields of *H= 500, 2000* and *10000 Oe*, using zero-field-cooled (ZFC), field-cooled (FC), and thermo-remnant protocols in the temperature regime between *50 K* and *300 K*. Magnetic hysteresis loops were recorded at *T=50, 220* and 300 K. An Agilent HP4194A-350 frequency-analyzer was used in order to determine the relative dielectric constant in the frequency regime from *100 Hz* to *40 MHz*. The hysteretic behaviour of electrical polarization as a function of applied electric field was measured with a Radiant Technologies ferroelectric tester, on the application up to *20,0 kV/cm*. The eventual appearance of an energy gap at room temperature was examined by using a VARIAN Cary 5000 UV–Vis–NIR diffuse reflectance spectrophotometer, which has an integration sphere with a PMT/Pbs detector.

# 3 Theoretical Calculations

The calculation of band- and electronic-structure for the complex perovskite $LaBiFe_2O_6$ can be seen as a many body problem of ions and electrons. The calculations are performed by employing the Full-Potential Linearized Augmented-Plane Wave (FP-LAPW) method, in the framework of Density Functional Theory (DFT) and implemented in the WIEN2k code [23,24]. The FP-LAPW consists in the calculation of solutions for the Kohn-Sham equations by the first principles method. In the calculations reported here, we use a parameter $RMTK_{max}=8$, which determines matrix size (convergence), where $K_{max}$ is the plane wave cut-off and RMT is the smallest of all atomic sphere radii. We have chosen the muffin-tin radii for La, Bi, Fe and O to be *2.50, 2.50, 1.89*, and *1.63*, respectively. The exchange and correlation effects were treated by using the Generalized Gradient Approximation (GGA) [25]. This potential considers the difference between the electronic densities for the two distinct spin orientations from the beginning. The density of states is calculated by the histogram method, and the position of the Fermi level is found by integrating over the density of states for both up and down spin orientations. The self-consistent calculations are considered to be convergent when the total energies of two successive iterations agreed to within $10^{-4}$ *Ry*. We adjusted the Fermi energy to zero. The integrals over the irreducible Brillouin zone are performed up to *140 k*-points.

# 4 Results and Discussion

### 4.1. Crystal Structure

Systematic monitoring of structural changes during the synthesis process of the $LaBiFe_2O_6$ material was carried out by XRD. In Figure 1 the diffraction patterns for diverse temperature treatments are presented. Figure 1a corresponds to the mixture of precursor powders $La_2O_3$, $Bi_2O_3$ and $Fe_3O_4$ after being macerated in an agate mortar at room temperature.

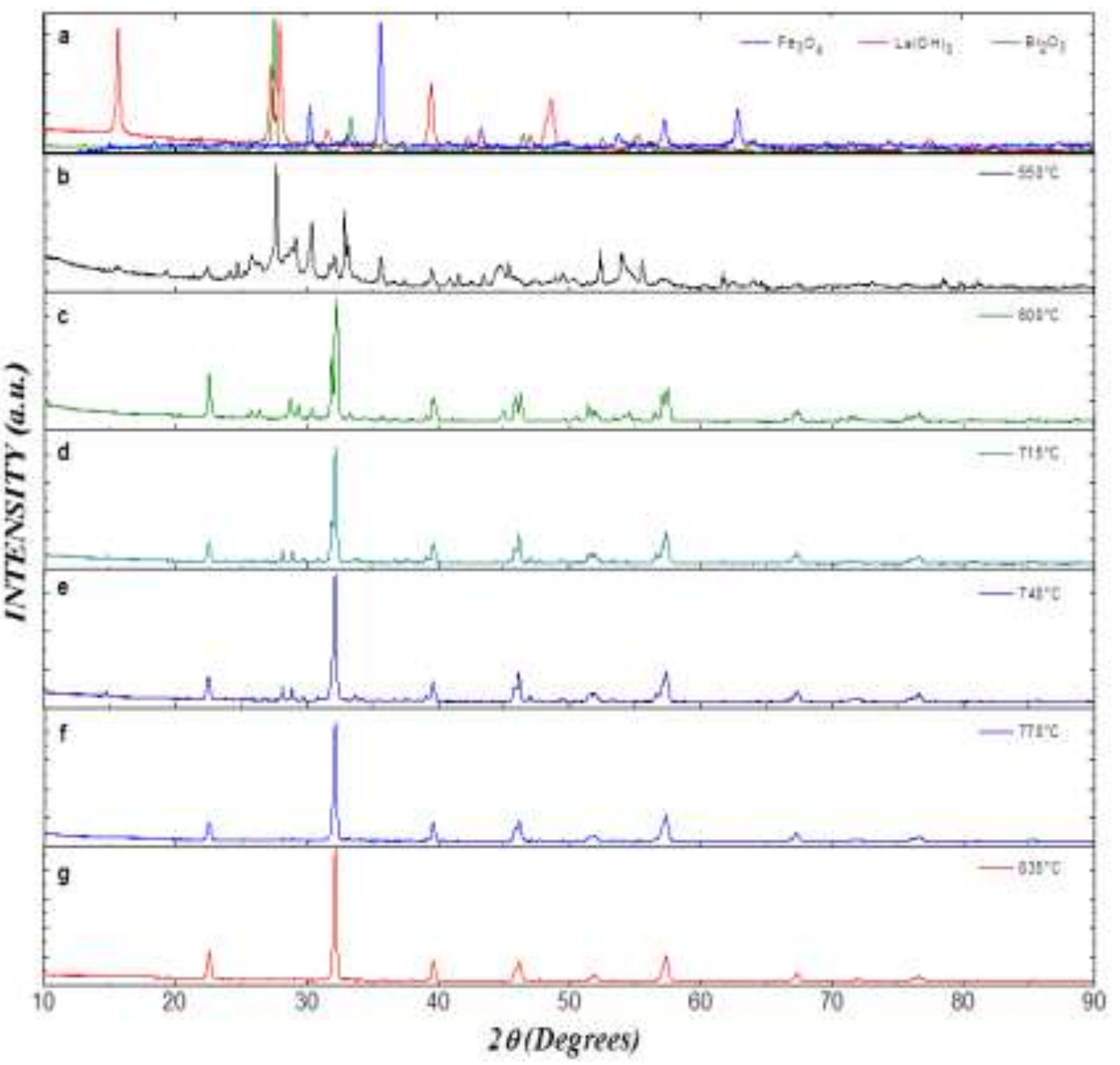

Figure 1. XRD patterns exemplifying the thermal process applied to the production of the $LaBiFe_2O_6$ samples.

From the figure 1a it was found that the $La(OH)_3$ crystallizes in a hexagonal structure, *$P6_3/m$ (#176)* [26], $Bi_2O_3$ in a monoclinic cell, *$P2_1/c$ (#14)* space group [27], and $Fe_3O_4$ in a cubic structure, *Fd-3m (#227)* space group, as reported in the ISCD cards [28]. Figures 1b to 1g represent the diffractograms obtained after heat treatments at *550 $^oC$, 600 $^oC$, 715 $^oC$, 740 $^oC$, 770 $^oC$,* and *835 $^oC$*, respectively. Figure 1b permits to infer that a reaction is taking place, as some reflections associated with precursor powders have disappeared and new ones begin to appear. In 1c, it is observed that at *600 $^oC$* a majority crystallographic phase begins to form but strong multiphase evidence there is still due to the presence of $Bi_2O_3$ and $La_2O_3$ precursors oxides, represented by the diffraction peaks that appear at the $26^o < 2\theta < 30^o$.

From the thermal process at temperature *600 $^oC$* (figure 1c) up to treatment at *835 $^oC$* (figure 1g), the intensity of peaks corresponding to the binary oxides $Bi_2O_3$ and $La_2O_3$ have disappear and a single crystallographic phase is reached.

Result of refinement of the XRD pattern corresponding to the heat treatment exemplified in figure 1g is show in figure 2. In this picture, the experimental data is represented by symbols, the simulated diffractogram is drawing as a red line, the background is indicated through a green line, and the difference between the experimental and theoretical patterns by a blue line.

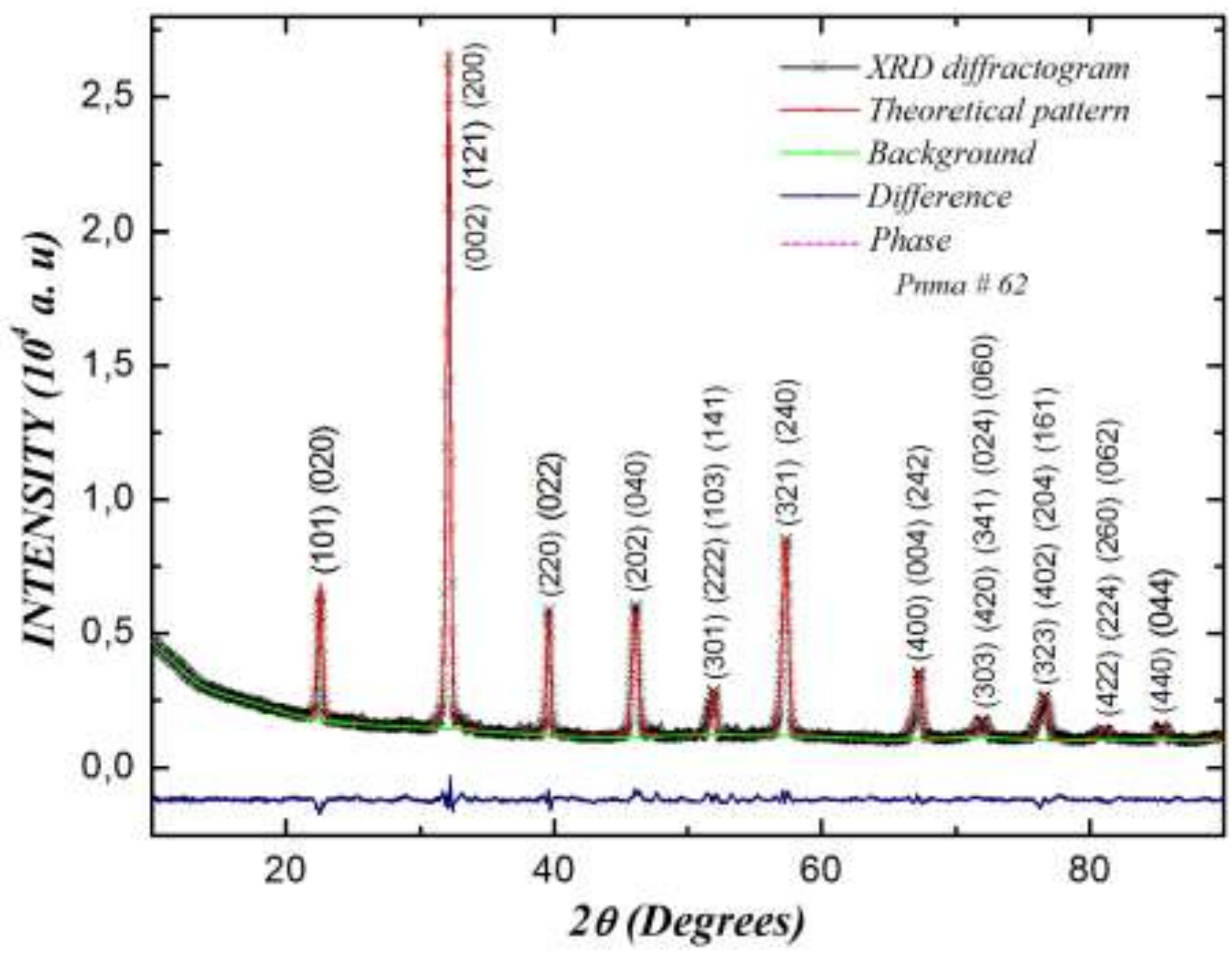

Figure 2. Refined XRD pattern including the diffracting planes associated to the experimental reflections for the $LaBiFe_2O_6$ material.

The respective diffraction planes appear indexed above the peaks of the pattern. The cell parameters obtained from the refinement are presented in table 1, where *a, b* and *c* represent the lattice parameters; *x, y* and *z* are the atomic positions of atoms in the unit cell; *α, β* and *γ* are the angles between the cell coordinates; $\chi^2$; $R(F^2)$; $R_p$ and $R_{wp}$ are the reliability parameters, which give information about the quality of the refinement procedure. $LaBiFe_2O_6$ crystallizes in a perovskite material with orthorhombic structure, *Pnma (#62)* space group with parameter $b > a, c$, as shown in figure 3. Although it is expected that the *Pnma* space group corresponds to simple perovskites, there are reports of complex perovskites which crystallize in structures belonging to this space group [29,30].

As observed in figure 3 and from the Wyckoff positions explicit in table 1, it is clear that the $La^{3+}$ and $Bi^{3+}$ cations share the same atomic position (with or without cationic ordering). This probably occurs because of the similarity between the ionic radii $r_{La}^{3+}=1,22$ Å and $r_{Bi}^{3+}=1,24$ Å. In the ideal case, the *Pnma* group is identified by the Glazer tilt system $a^+b^-b^-$, which is characterized by two angles of rotation with quite similar magnitude and opposite directions [31].

Table I. Cell parameters of the $LaBiFe_2O_6$ complex perovskite. The reliability factors are $\chi^2$= *2,8650; $R(F^2)$=5,68%; $R_p$=2,93 %* and *$R_{wp}$=3,95 %.*

| | | *Atomic coordinates* | | ***Pnma Space Group*** |
|---|---|---|---|---|
| ***Atom*** | ***x*** | ***y*** | ***z*** | ***Cell parameters*** |
| La, Bi | 0,50160 | 0,25000 | 0,49479 | *a=5,5961(1) Å* |
| Fe | 0,00000 | 0,00000 | 0,50000 | *b=7,8444(2) Å* |
| $O_1$ | 1,08218 | 0,25000 | 0,58605 | *c=5,5583(2) Å* |
| $O_2$ | 0,25984 | 0,01053 | 0,78898 | *α=β=γ=90.0°* |

The crystallographic angle $\beta=90^o$ obtained for this structure is in agreement with the expected value for an orthorhombic structural group. The Glazer tilt system suggests the occurrence of a cell size relative to an aristotype cubic unit cell with parameter $a_p$, such that the network parameters in terms of $a_p$ are given by $a\approx\sqrt{2}a_p$, $b\approx 2a_p$ and $c\approx\sqrt{2}a_p$ [32]. As observed in figure 3a, out of phase tilting of the $FeO_6$ octahedra occurs along the *b* and *c* axis. As will be seen later, the dramatic octahedral distortion observed in Figure 3b, could be responsible for the unconventional electric and magnetic responses of the $LaBiFe_2O_6$ material.

Likewise, 3c and 3d clearly show that octahedra are not perfectly aligned along the crystallographic axes, presenting rotation and inclinations around these axes, such that the distances of the bonds between $Fe^{3+}$ cations and the $O^{2-}$ anions are different along the three sub-axes of each octahedron. Thus, the distance Fe-O(1) is *2,07 Å*, whereas Fe-O(2) assumes two values: *1,79 Å* and *2,17 Å*. Another evidence of distortion is the variety of interatomic distances between cations La, Bi and the anions O, as is shown in table 2. As can be clearly seen in Table 2, there is a marked difference between the distances for several Bi,La-O(1) and Bi,La-O(2) bonds. These differences have to do with the octahedral distortions, which are responsible for the displacement of oxygen anions from their ideal positions of high symmetry.

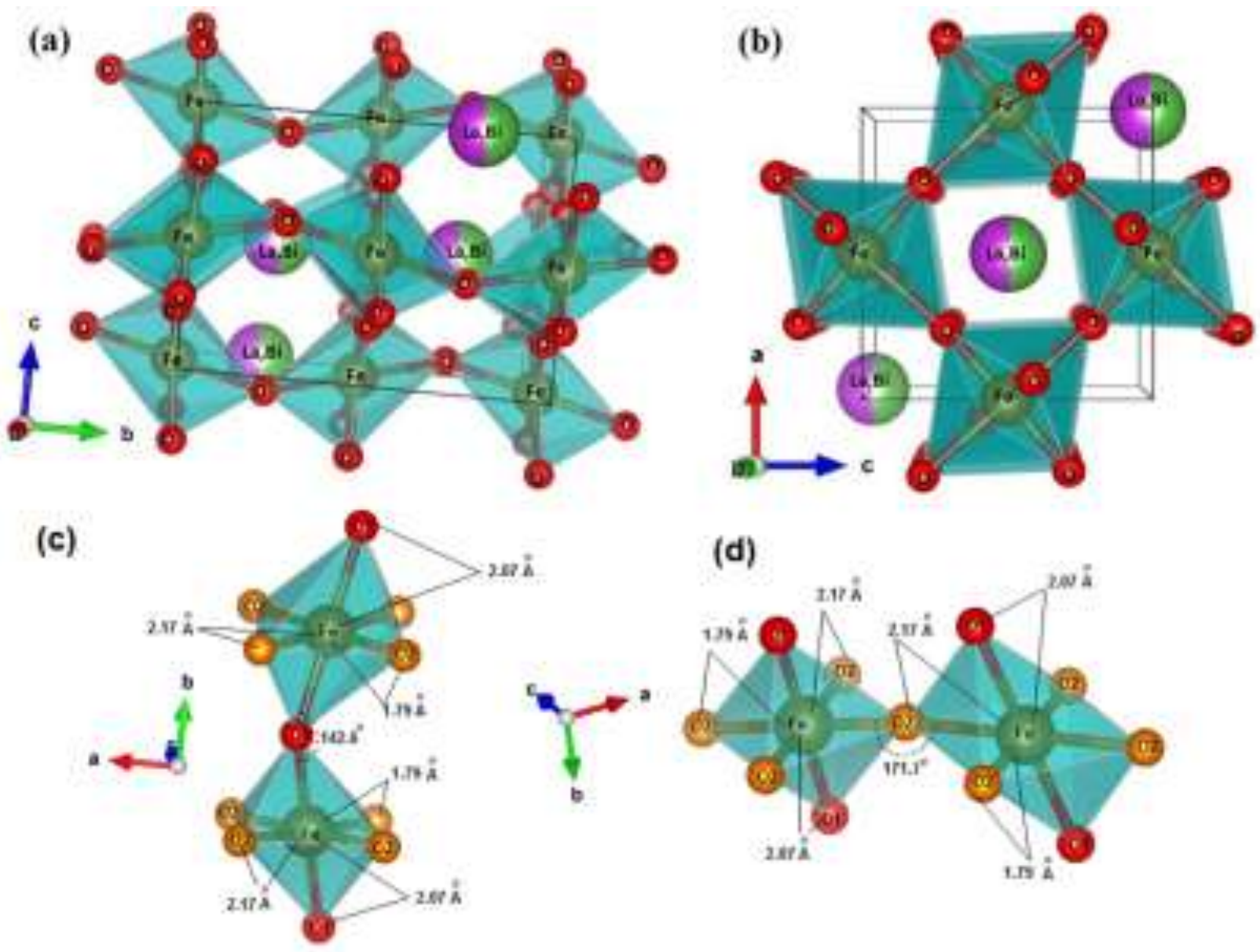

Figure 3. Structure of the $LaBiFe_2O_6$ complex perovskite constructed from the data obtained by Rietveld refinement of the experimental diffraction pattern; a) view along the *a*-axis (*b*-*c* plane), b) view along the *b*-axis (*c*-*a* plane), c) angle and atomic distances view in the *a*-*c* plane, and d) in 3D.

The differences between the interatomic distances, and rotations and inclinations of the octahedra, clearly demonstrate the inhomogeneity of the octahedral distortions which has place within the crystal structure of the material.

In order to determine the mean size of the crystalline domains we apply the Scherrer formula, which is given by

$$\beta_{hkl} = \frac{\kappa\lambda}{FWHM \cos\theta}, \tag{2}$$

where, $\beta_{hkl}$ represents the crystallite size and depends only on the *hkl* direction; $\kappa$ is a dimensionless shape factor (typical value *0,9*); $\lambda = 1{,}54056$ *Å* is the XRD wavelength, FWHM is the line broadening at half the maximum intensity and θ is the Bragg angle.

Table II. Interatomic distances of the Bi, La-O bonds for the specific atomic sites of the O ion and the respective diffraction planes.

| La,Bi-O(1) Distance (Å) | O(1) Ion Sites | XRD Planes |
|---|---|---|
| 4,5569 | *-x+ ½, -y, -z+ ½* | (1 1 -1) |
| 3,2598 | *x+ ½, y, -z+ ½* | (-1 0 0) |
| 4,5569 | *-x+ ½, -y, z+ ½,* | (1 0 -1) |
| 5,1243 | *-x, -y, -z* | (1 1 1) |
| 2,4014 | *x, y, z* | (-1 0 0) |
| 5,1243 | *-x, -y, -z* | (1 0 1) |
| **La,Bi-O(2) Distance (Å)** | **O(2) Ion Sites** | **XRD Planes** |
| 4,7121 | *x+ ½, -y+ ½, -z+ ½* | (1 0 1) |
| 2,8343 | *x, -y+ ½, z* | (0 0 0) |
| 2,7615 | *-x+ ½, y+ ½, z+ ½,* | (0 0 -1) |
| 2,9063 | *-x, y+ ½, -z* | (1 0 1) |
| 4,9594 | *x+ ½, -y+ ½, -z+ ½* | (0 0 0) |
| 4,5553 | *x, -y+ ½, z* | (0 0 -1) |

The obtained crystallite mean size was $\beta_{hkl}=54,0 \pm 7,4$ *nm*. There is another way to get the mean size of crystallite through the Rietveld refinement by considering the Debye Scherrer cone in the scattering angle $2\theta$. The GSAS code applies the equation

$$p = \frac{1800\kappa\lambda}{\pi X'} , \tag{3}$$

where, κ represents the Scherrer constant, $\lambda=1,54056$ *Å* is the XRD wavelength and $X'$ is the Lorentzian coefficient given by $X+P_{tec}$ (Lorentzian coefficient of particle size and anisotropy factor) for the parallel crystallite size and $X$ for the perpendicular crystallite size. Total crystallite size is calculated as the average of the parallel and perpendicular sizes. The mean size of crystallite given by the GSAS code is $p=56,4\pm5,1$ *nm*. This value is very similar to that which was found by directly applying the Scherrer formula.

### 4.2. Morphology

Figure 4 exemplifies granular characteristics of the samples obtained from SEM measurements with a magnification of 32500X. As observed in figure 4a, there are two types of grains, whose sizes are studied by means the profile presented in figure 4b. A first group of grains of sizes reaching *1,096±0,034 μm* and are strongly disseminated each other is clearly perceived which is identified in the profile as agglomerate grains. These are composed by small grains with sizes very close to the second type of grains observed, which reaches *0,309±0,018 μm*. These two types of grains have been clearly classified based on the analysis presented in Figure 4b. Our results reveal that the group of small grains is a residue due to heat treatment, where the small grains are formed but are not enough disseminated to contribute to the formation of larger grains.

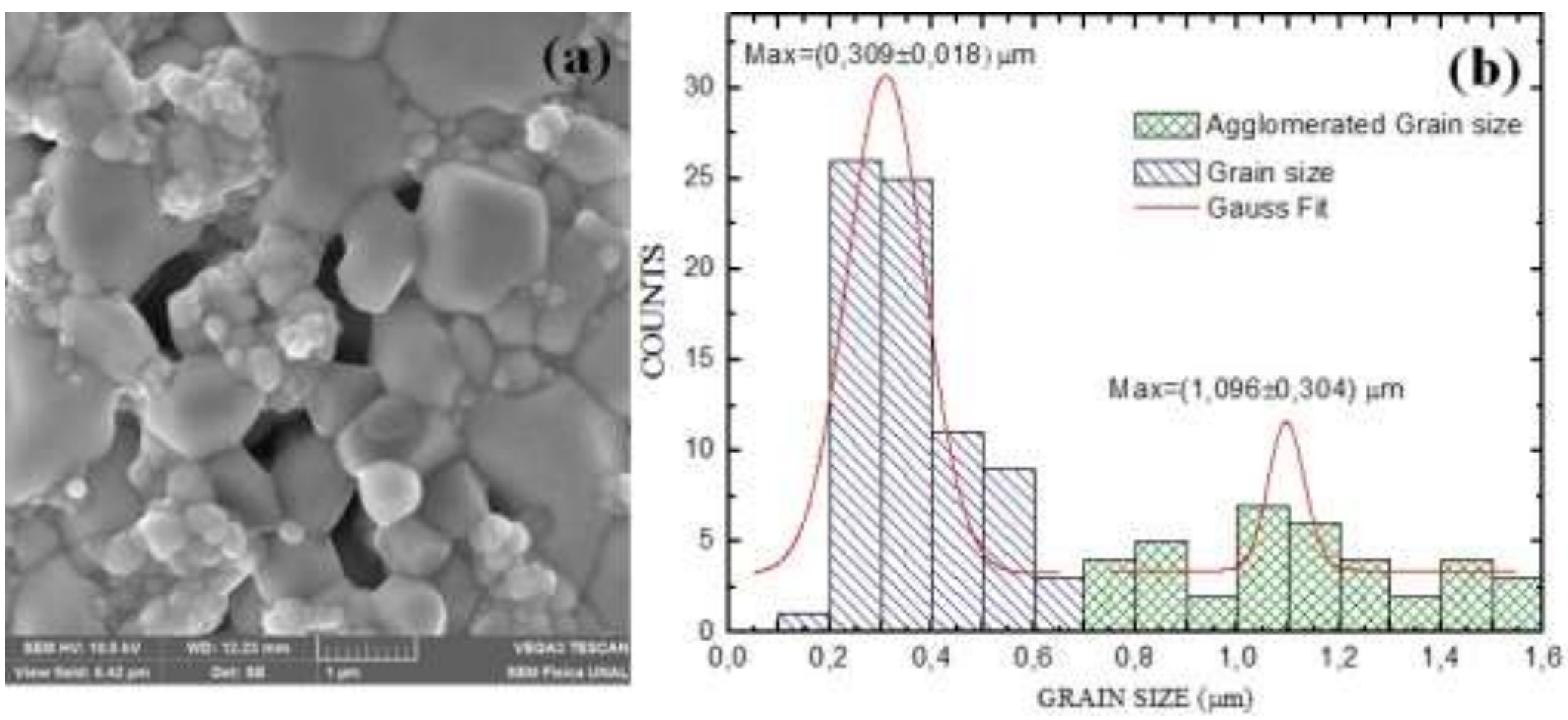

Figure 4. SEM image of the granular surface of the $LaBiFe_2O_6$ samples (a) and analysis of the mean grain size (b).

By using the microprobe coupled to scanning electron microscope, EDS spectra from electron beams incident on the different surface grains of the samples (see Figure 5), allowed to establish that they contain the stoichiometry of the expected composition up in *97 %*, according to the values specified in table 3.

Table III. EDX experimental and expected weight percentages for each atom in the $LaBiFe_2O_6$ material.

| ***Atom*** | ***Exper. (w%)*** | ***Theor. (w%)*** |
|---|---|---|
| *Bi* | 38,05 | 37,616 |
| *La* | 25,95 | 25,003 |
| *Fe* | 19,64 | 20,100 |
| *O* | 16,36 | 17,281 |

From the results of table 3 and figure 5, it can be stated that samples synthesize a single material whose stoichiometry is close to the expected, thus the conclusion obtained through the analysis of the experimental data of XRD is reinforced: a single crystallographic phase is obtained. In the table, experimental percentages which correspond to the mass of the different elements of the material (column *Exper. w%*) there are tabulated and compared with the percentages calculated from the nominal composition of the material (column *Theor. w%*).

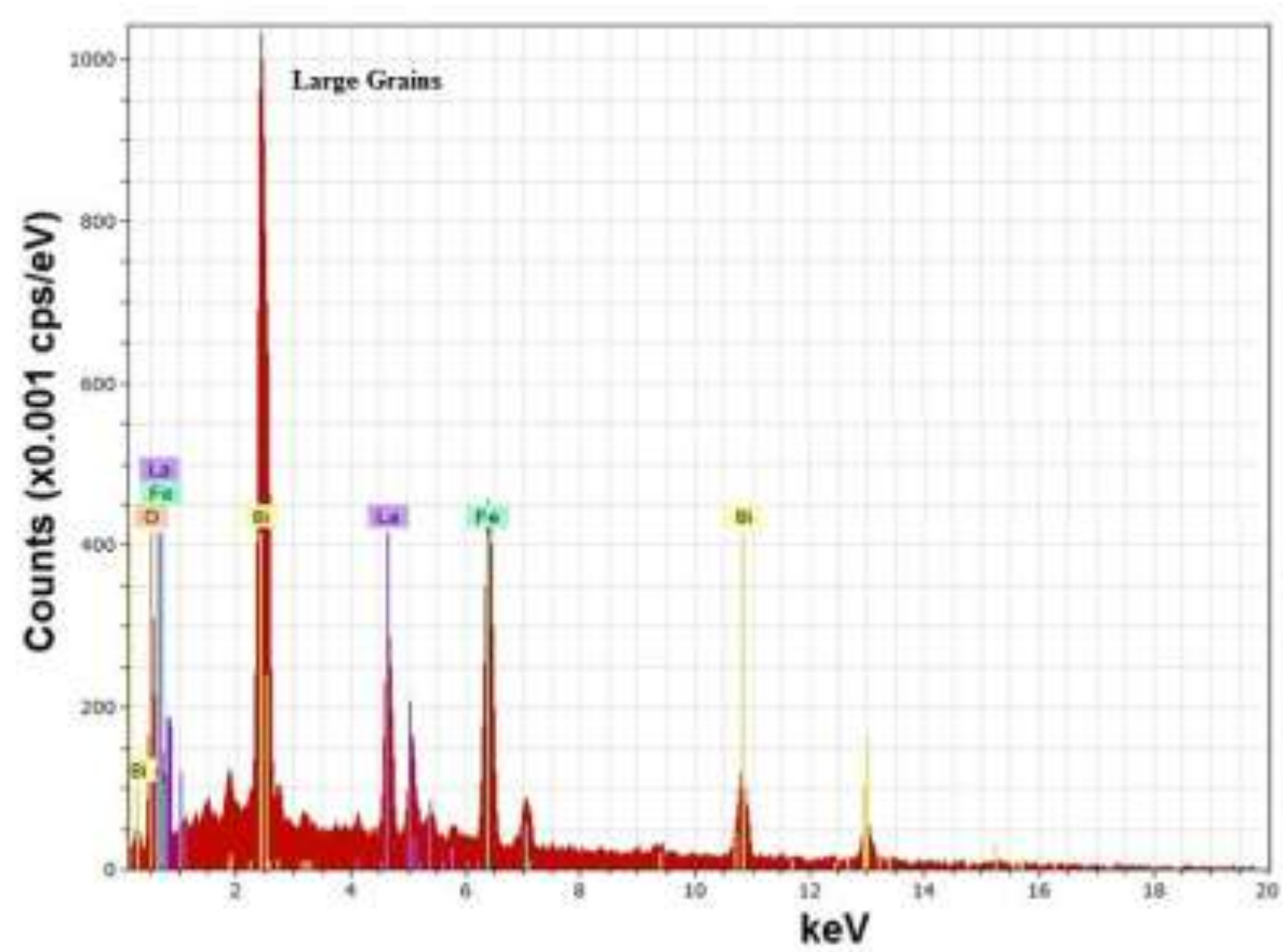

Figure 5. EDX spectrum obtained when the electron beam impinges on the large grains of the material. The spectrum is virtually identical for the small grains.

### 4.3. Electric Polarization and Dielectric Response

In order to study the polarization of the material for several applied electric fields, hysteresis curves as those presented in Figure 6a were experimentally obtained on the applications of a signal with voltages up to *2 kV* and period of *0,01 s*. Values of characteristic parameters are reported in table 4, where $E_C$ are the coercive fields, $P_R$ represents the remnant polarization and $P_S$ denotes the saturation polarizations for each hysteretic curve. The hysteresis loops of figure 6a evidence a typical behaviour of strong dielectric loss [33].

Table IV. Saturation polarizations, remnant polarizations and coercive fields obtained for hysteresis polarization curves as a function of several applied electric fields.

| ***E*** (kV/cm) | $P_S$ (x$10^{-2}$ $\mu C/cm^2$) | $P_R$ (x$10^{-2}$ $\mu C/cm^2$) | $E_C$ (kV/cm) |
|---|---|---|---|
| 1,00 | 0,701 | 0,317 | 0,365 |
| 2,00 | 1,398 | 0,633 | 0,732 |
| 4,00 | 3,112 | 2,053 | 2,547 |
| 6,00 | 4,289 | 3,081 | 3,820 |
| 8,00 | 5,802 | 4,272 | 5,102 |
| 10,0 | 7,097 | 5,340 | 6,378 |

From the values of $P_S$ and the corresponding electric fields taken from figure 6a, the graph of saturation polarization as a function of applied fields was prepared as shown in figure 6b.

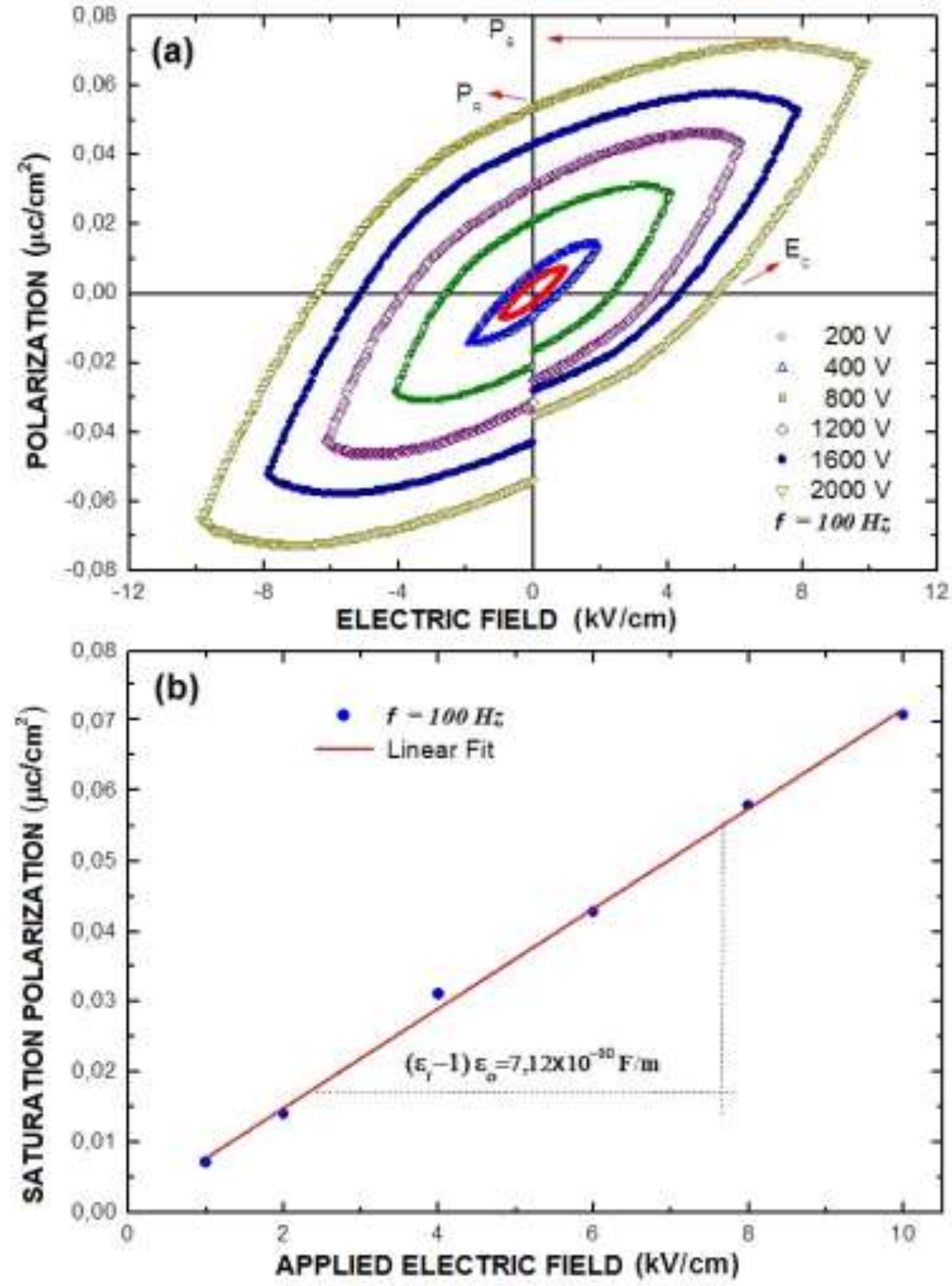

Figure 6. Hysteretic response of the electric polarization on the application of different electric fields.

Using the linear fit of the experimental data the value $(\varepsilon_r\text{-}1)\varepsilon_o=7,12\text{x}10^{-10}$ *F/m* was obtained and the relative dielectric constant was determined to be $\varepsilon_r=81,45$. Results of relative dielectric constant measurements are show in figure 7. It is observed that for a frequency of *100 Hz* the dielectric constant has a value $\varepsilon_r=81,53$, which is strongly close to that value obtained from the saturation polarization in the hysteresis curves of polarization as a function of applied fields. On the other hand, when frequency is increased, the $\varepsilon_r$ value diminishes drastically in the frequency range between *100 Hz* and *500 Hz* and then decreasing smoothly with minimum values of *13,05* for frequencies up to *40 MHz*. This value is close to the relative dielectric constant reported for $BiFeO_3$ single crystals ($\varepsilon_r=83$) [34] but relatively low as compared with the colossal dielectric response reported for $LaFeO_3$ [35]. Also, the dielectric constant at high frequencies is close to the expected value for extrinsic semiconductors [36].

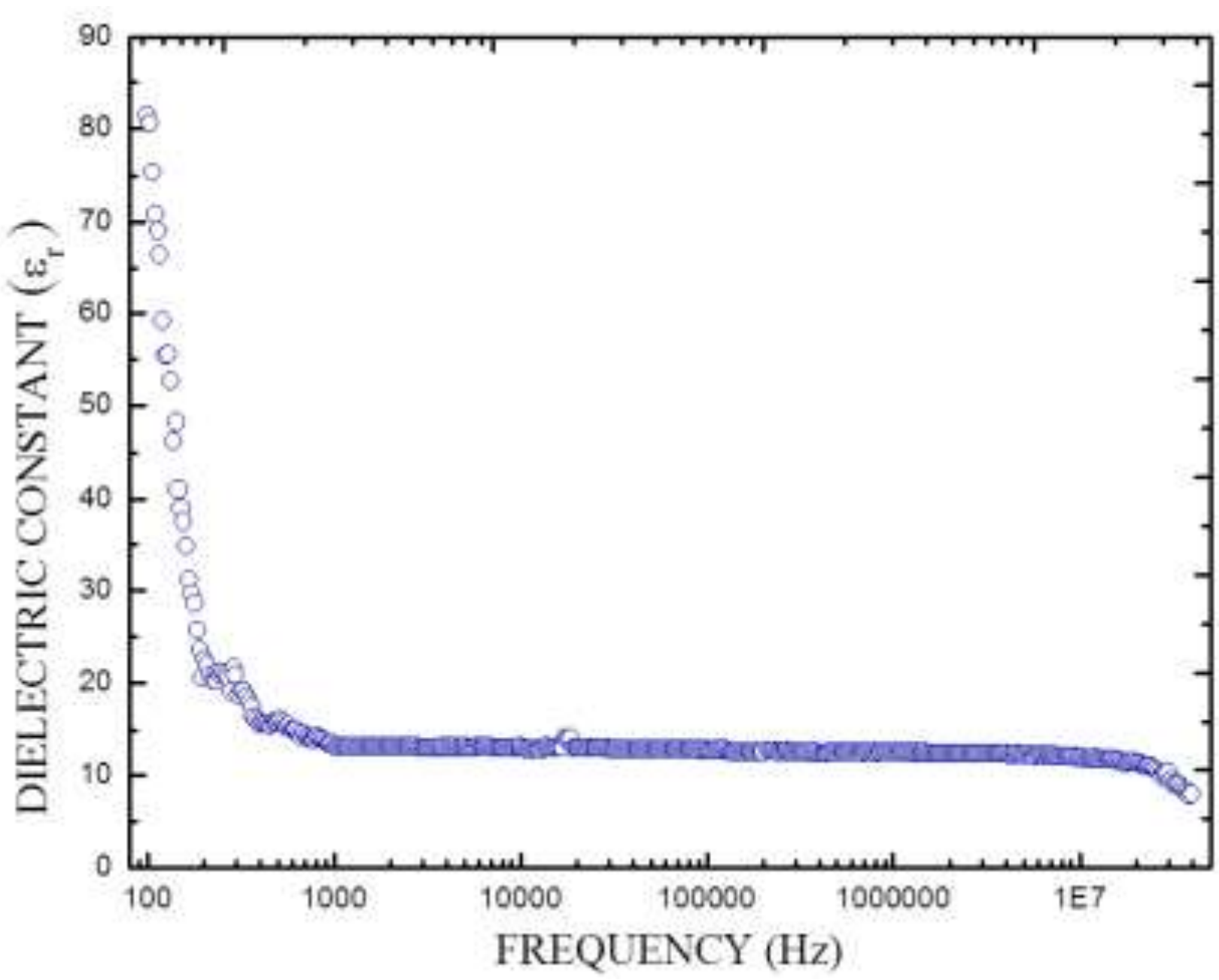

Figure 7. Dielectric constant as a function of frequency measured for the $LaBiFe_2O_6$ perovskite.

### 4.4. Magnetic Susceptibility and Magnetization

The magnetic susceptibility response as a function of temperature on the application of intensity fields H=*500 Oe, 2 kOe* and *10 kOe* is exemplified in figure 8. Measures were performed following the Zero Field Cooling (ZFC) procedure, cooling in the absence of magnetic field, applying the field at low temperatures and measuring the susceptibility during heating of the sample, and in the Field Cooled (FC) recipe, measuring it in the presence of the applied field while the sample is cooled. In perfectly ordered materials it could be expected that the ZFC and FC susceptibility responses were identical, except in the case of so-called spin glasses [37]. In our case, we would expect the $La^{3+}$ and $Bi^{3+}$ cations maintain a spatially ordered arrangement within the structure of the $LaBiFe_2O_6$ complex perovskite, alternating along the sub cells constituting crystallographic unit cell. Meanwhile, this type of cation ordering was not specifically concluded from the Rietveld analysis of the XRD patterns.

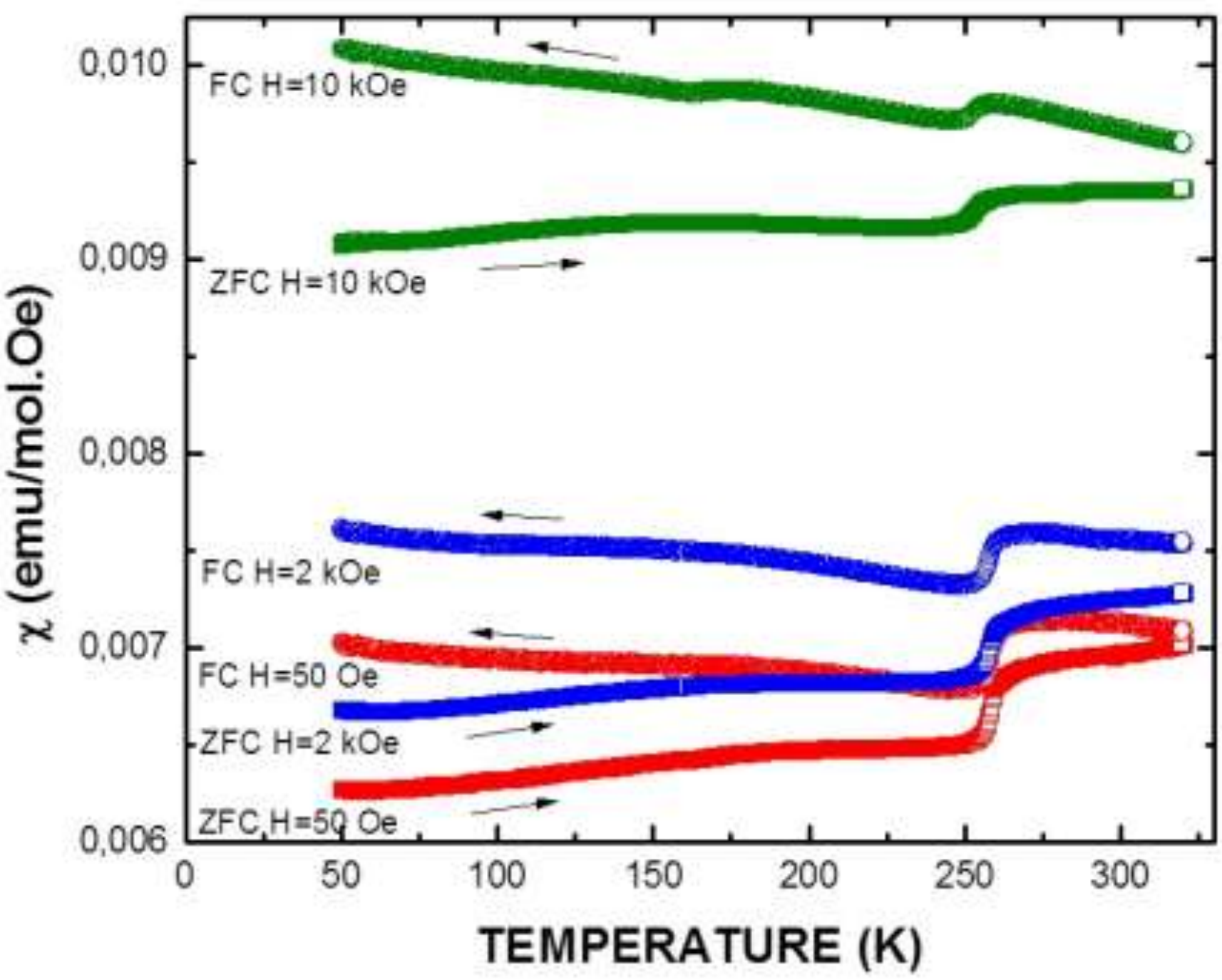

Figure 8. Magnetic susceptibility as a function of temperature for the $LaBiFe_2O_6$ material on the application of *H=500 Oe, 2 kOe* and *10 kOe*.

Therefore, it is possible that the cations $La^{3+}$ and $Bi^{3+}$ are not perfectly ordered along the structure, constituting an outline of cationic disorder which could cause complex distortions in the $FeO_6$ octahedra that affect the magnetic exchange, resulting in responses of the frustrated type. Thus, the inhomogeneity of the interactions between the magnetic moments and competition between them may result in a great randomness of the magnetic spins of the material, giving rise to a marked difference between ZFC curves and FC susceptibility, as observed in figure 8. Moreover, measurements in the presence of different magnetic fields show the ferromagnetic character of the perovskite $LaBiFe_2O_6$ in the temperature regime $50\ K<T<300\ K$. However, in the temperature range of functionality of our equipment, it was not possible to determine the value of the Curie temperature for this material.

A relevant characteristic in figure 8 is the occurrence of an apparent abnormal transition in *T=258 K*. The transient drop in magnetic susceptibility at this temperature value for all applied fields may be related to magnetocrystalline anisotropy effects. As the temperature decreases and approaches the value of *T=258 K*, the structural distortions as rotations and inclinations of octahedra, as well as differences of interatomic distances, introduce contributions that are not on the axis of easy magnetization. These are evident through a sharp drop in the susceptibility of the material, usually accompanied by a decrease in the saturation magnetization, as observed in figure 9.

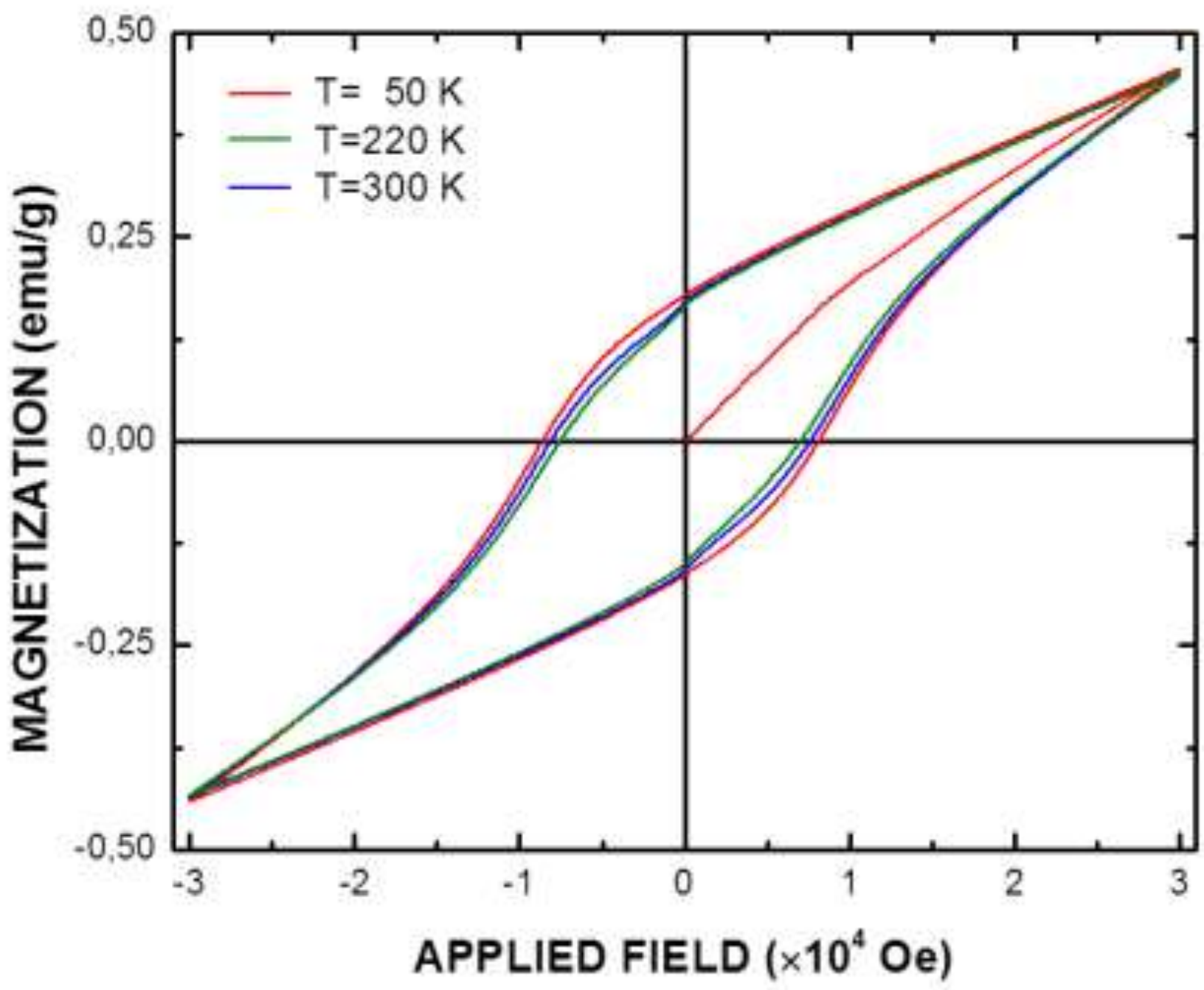

Figure 9. Magnetization as a function of applied field measured at the temperatures *T=50 K, 220 K* and *300 K*.

In this picture, a ferromagnetic behaviour of magnetization measurements as a function of applied field at *T=50 K, 220 K* and *300 K* is observed. It is clear that the higher saturation magnetization occurs at *T=50 K*; at a temperature of *220 K* the saturation magnetization is lower and then increases slightly for *T=300 K*. Another cause of this anomalous behaviour has to do with the shape anisotropy due to the inhomogeneous nature of the polycrystalline grains of the system, as shown in Figure 4. The presence of sub-micrometric grain-size can introduce border and surface effects, as well as the possibility of formation of independent magnetic domains which can behave as a superparamagnetic sub-material, giving rise to the apparent linearity of the saturation magnetization regime in the hysteretic curves of figure 9. In the same form, these domains can make an important contribution to the frustrated nature of the system which is evident in the irreversibility of the susceptibility curves (ZFC and FC) shown in Figure 8. A third cause of this reentrancy in magnetic susceptibility can be related to stress anisotropy. The results of electrical polarization (Figure 6) show that at room temperature the $LaBiFe_2O_6$ material behaves like a ferromagnetic with a small degree of dielectric loss. In the same way, the results of magnetization (Figure 9) suggest a ferromagnetic response of this ceramic compound at room temperature. Meanwhile, a supposed coexistence between ferromagnetism and ferroelectricity not guarantee coupling between the ferroic order-parameters in the material. But this coupling may be associated with structural spontaneous deformation effects at $T=258\ K$, producing a sudden decrease in the magnetic susceptibility due to the energy cost required for the coupling. Thus, the stress anisotropy can occur due to the magnetoelastic coupling, which is also related with crystalline anisotropy and inhomogeneity of the material. All the above possibilities may be summarized in the so called Hopkinson effect, which is characterized by an abrupt decrease of magnetic susceptibility below the Curie temperature [38]. The remnant magnetization and coercive field measured correspond to values of $M_R=0,177$ *emu/g* and $H_C=8,3\ kOe$ respectively.

### 4.5. Diffuse Reflectance

In order to experimentally estimate the value of energy gap, Ultra-Violet Visible (UV-Vis) Diffuse Reflectance measurements were carried out. Results are shown in figure 10. Four peaks are clearly observed in figure 10a for wave lengths $\lambda_1=313\ nm$, $\lambda_2=796\ nm$, $\lambda_3=1400\text{-}1600\ nm$ and $\lambda_4=2213\ nm$. The peak of longer wavelength ($\lambda_4$), i.e., less energy, is related with effects introduced by the experimental technique, since it detects vibration modes in the structure due to movement of the entire cell in addition to the vibration of isolated atoms. The other three peaks observed ($\lambda_1$, $\lambda_2$ and $\lambda_3$), corresponding to energies *3,96 eV, 1,56 eV* and *0,83 eV* (see Figure 10b) are related with electronic excitations due to electronic transitions from valence band to the conduction bands. The presence of these three peaks in the spectrum is expected for the $LaBiFe_2O_6$ material due to its structural characteristics, because the symmetry of its atomic positions in the *Pnma (#62)* space group suggests the transitions $O_{2p}$-$Bi_{6p}$ (*0,83 eV*), $O_{2p}$-$Fe_{3d}$ (*1,56 eV*) and $O_{2p}$-$La_{5d}$ (*3,96 eV*) [39] on the irreducible representation of vibrations given by

$$\Gamma = 7A_g + 7B_{1g} + 5B_{1g} + 5B_{2g} + 8A_u + 7B_{1u} + 9B_{2u} + 9B_{3u}, \qquad (4)$$

where, the first five terms correspond to vibration modes which are observable by Raman spectroscopy, and the last three modes appear as peaks measured using less energy spectroscopy, such as infrared.

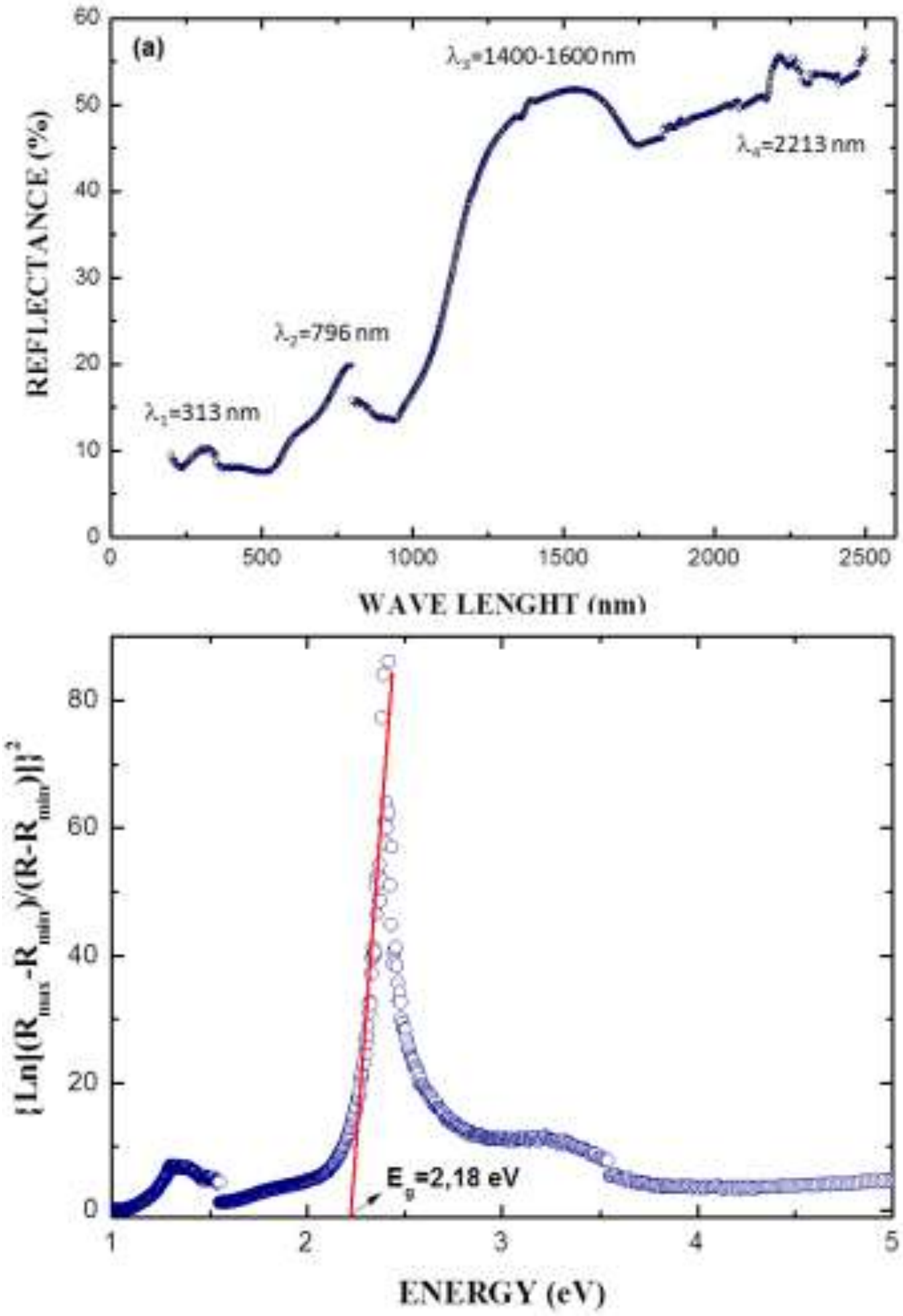

Figure 10. Diffuse reflectance measurements in the $LaBiFe_2O_6$ samples, a) % reflectance as a function of wave length and b) Kubelka-Munk analysis.

An interesting detail in the spectrum of Figure 10a is the widening of the peak corresponding to $\lambda_3$ (from *1400 nm* up to *1600 nm*), which could be attributed to surface reflection effects as a result of the polycrystalline nature of the solid. On the other hand, a point of interest in the application of the Diffuse Reflectance technique in this work is the estimation of the energy gap of the material. For this purpose, the method of Kubelka-Munk type analysis [40] is applied in the way as implemented by V. Kumar et al [41].

The absorption coefficient is given by

$$\alpha_b = \frac{B\left(h\nu - E_g\right)^n}{h\nu}, \tag{5}$$

where, $h\nu$ is the absorbed energy, $B$ is the absorption constant, $E_g$ represents the gap energy and $n=1/2$ or $n=2$ if there are direct or indirect semiconductor transitions, respectively. The energy gap is equivalent to the absorbed energy when $\alpha_R = \ln R$, where $R$ is the measured reflectivity relative to the unity. By representing the $\left[\ln \frac{R_{\max} - R_{\min}}{R - R_{\min}}\right]^2$ vs $h\nu$, it is possible to determine the gap energy as showed in figure 10b. The results of diffuse reflectance are related to the chemical structure of the material. In the case of $Fe^{3+}$, the valence and conduction bands present a splitting of the *3d*-Fe levels in the crystal field as a result of the octahedral coordination of the $FeO_6$. From the analysis of the Diffuse Reflectance spectrum, the obtained band energy gap was *2,17 eV*, which corresponds to a semiconductor behaviour. The difference of this value as compared with reports for $LaFeO_3$ [42] and $BiFeO_3$ [43] is partially caused by the occurrence of three different $Fe_{3d}$-$O_{2p}$ bonds into the $FeO_6$ octahedra (*1,79 Å, 2,07 Å* and *2,17 Å*, as discussed in section 4.1). Likewise, the interatomic distances of the La-O and Bi-O bonds are not symmetrical because of the octahedral distortions.

### 4.6. Density of States and Band Structure

In order to determine the structural stability of the $LaBiFe_2O_6$ material for the *Pnma* (#62) space group obtained experimentally, a procedure of minimizing of the energy as a function of the volume of the unit cell was carried out. The obtained result was adjusted to the Murnaghan's state equation [44], which is obtained by assuming a linear behaviour of the compressibility modulus of a solid with respect to the pressure at constant temperature. The obtained curve is showed in figure 11.

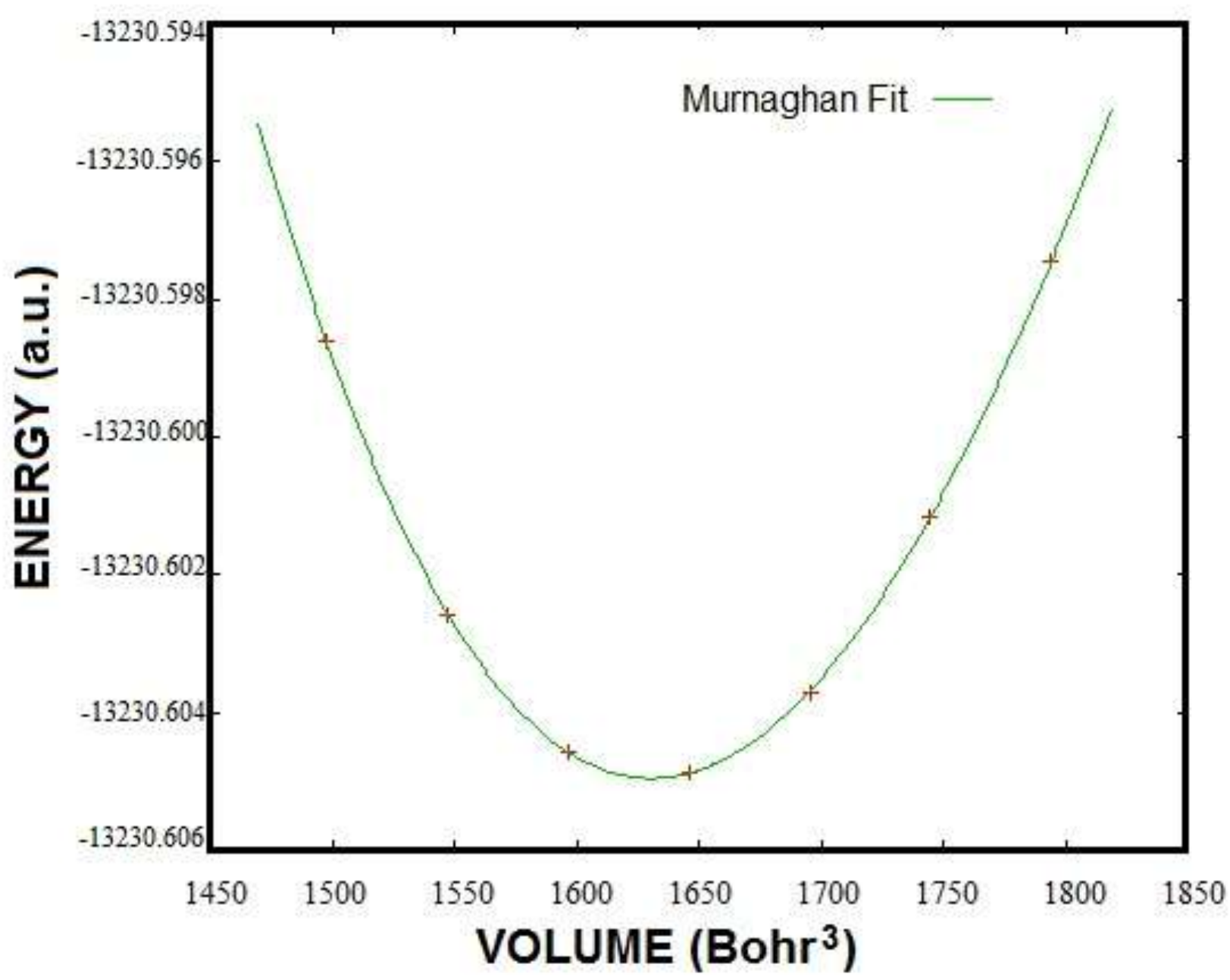

Figure 11. Curve of minimization of energy as a function of volume and fitting to the Murnaghan equation.

Thus, the volume obtained for the minimum energy corresponds to $V_o$*=1630 Bohr*$^3$, which is close to the experimental volume of the unit cell for the *Pnma (#62)* space group, $V_o$*=243,9985 Å*$^3$ *= 1646 Bohr*$^3$. Additionally, from the Murnaghan equation, the bulk modulus of the cell was calculated to be *153,02 GPa*.

In order to build the band diagram of the system, high symmetry points in the first Brillouin zone were selected. For the orthorhombic phase, *Pnma* (#62) space group, of the $LaBiFe_2O_6$ perovskite, *140 k* vectors in the irreducible zone were considered. The energy bands were calculated along the path Γ-X-S-Y-Γ-Z-U-R-T-Z|Y-T|U-X|S-R, as showed in figure 12.

Results of band structure for the up and down spin configurations are exemplified in figure 13. In this picture, the center of the first Brillouin zone is designated by the Greek letter Γ, located at the origin of coordinates system. The axis abscissa represents the points of high symmetry and the axis coordinate values of energy in eV.

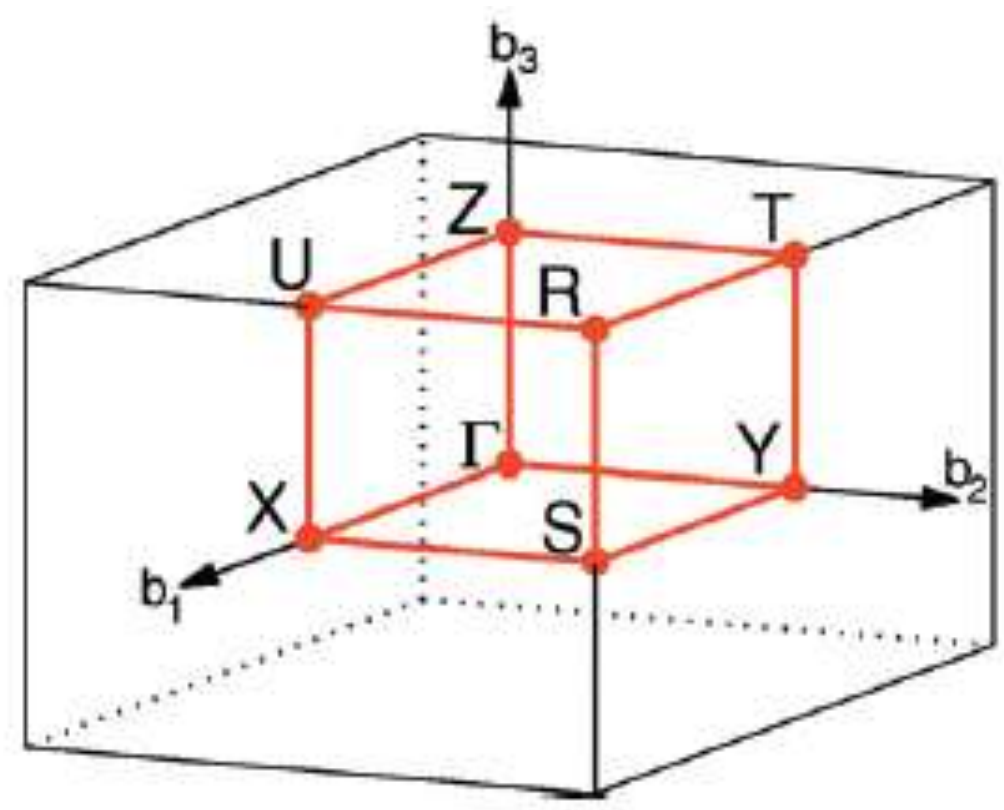

Figure 12. Path of calculation of the band energies in the first Brillouin zone of the orthorhombic perovskite structure.

Between two vertical lines there is a direction of the first Brillouin zone, along which the Hamiltonian is diagonalized in a collection of points. All energies corresponding to these points form the band structure, so that the diagram goes through the high symmetry directions of the first Brillouin zone, providing information about the evolution of energy with the wave vector. Fermi level corresponds to the zero energy. For the spin down configuration an energy gap $E_g=1,3\ eV$ is observed in the valence band.

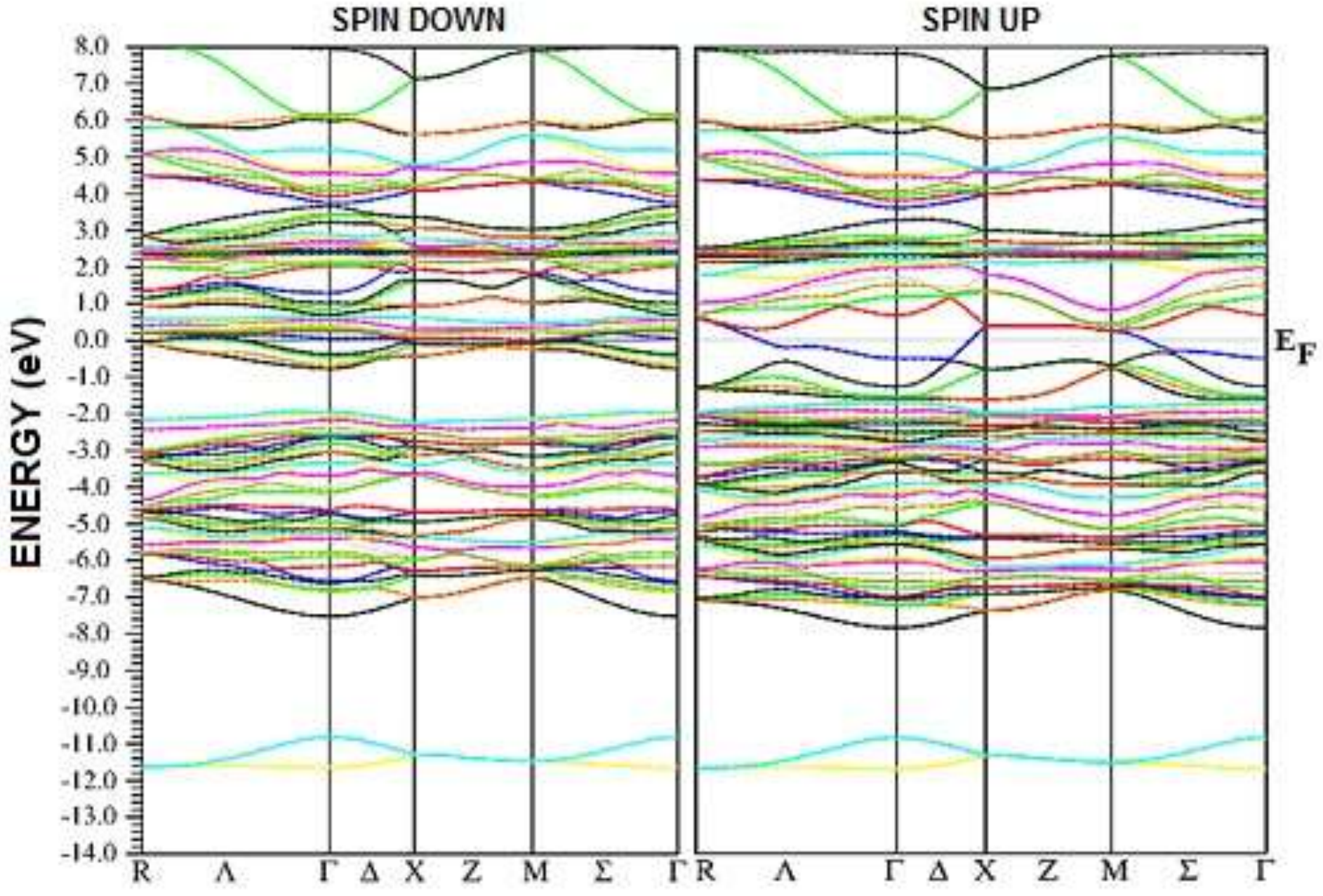

Figure 13. Band structures for the down and up spin orientations.

This value clearly suggests semiconductor behaviour. On the other hand, for the spin up configuration, no presence of energy gap due to the overlap of the bands caused by transitions, produced by the coulomb interaction between electrons and holes, which gives place to excitons. Then, in a narrowband semiconductor, excitons reduce the energy below the Fermi level. This state is also accompanied by deformations in the structural lattice. The value of minimum energy in the gamma direction is *-7,5 e*V to spin down orientation while for the spin up orientation is *-7,8 eV*, indicating a slightly higher hybridization of the *p* and *d* orbitals for the spin up configuration. Clearly there is a significant difference between the theoretical and experimental values of the gap. This dissimilar result has to do with two factors in the calculations by DFT: first, the relevance of the approximations introduced to the exchange energy and correlation, making this term is the most significant in the Hamiltonian, and second, the fact that the DFT does not predict the excited states because it calculates the energy in the ground state. On the other hand, it is known that in complex perovskites interactions between d orbitals of the cations and p orbitals of the anions introduce covalent hybridization terms [45]. This can be seen in Figure 14, where the densities of states of partial atoms La, Bi and Fe are superimposed, suggesting the presence of covalent bonds [46-47], especially for the spin down configuration. The calculated total effective magnetic moment was *13,7 eV*, which is relatively close to the experimental value *12,3 eV*, since the deviation is *≈10%.* The magnetic contribution is mainly due to the *d* orbitals of Fe, because the density of states of this atom is the closest to the Fermi level and higher value. These orbitals suffer a splitting in the conduction band to the spin up configuration, which can be seen in Figure 13 between the symmetry points Γ and X for the energy range between *-1,0 eV* and *1,0 eV*. This splitting has place as a consequence of the crystalline field, with two spin up electrons in the $e_g$ states $d_{x^2-y^2}$ and $d_{z^2}$, called high spin configuration, which contribute to the conductor behaviour in the density of states for the spin down polarization, and four spin up electrons distributed in the $t_{2g}$ states $d_{xy}$, $d_{xz}$ and $d_{yz}$.

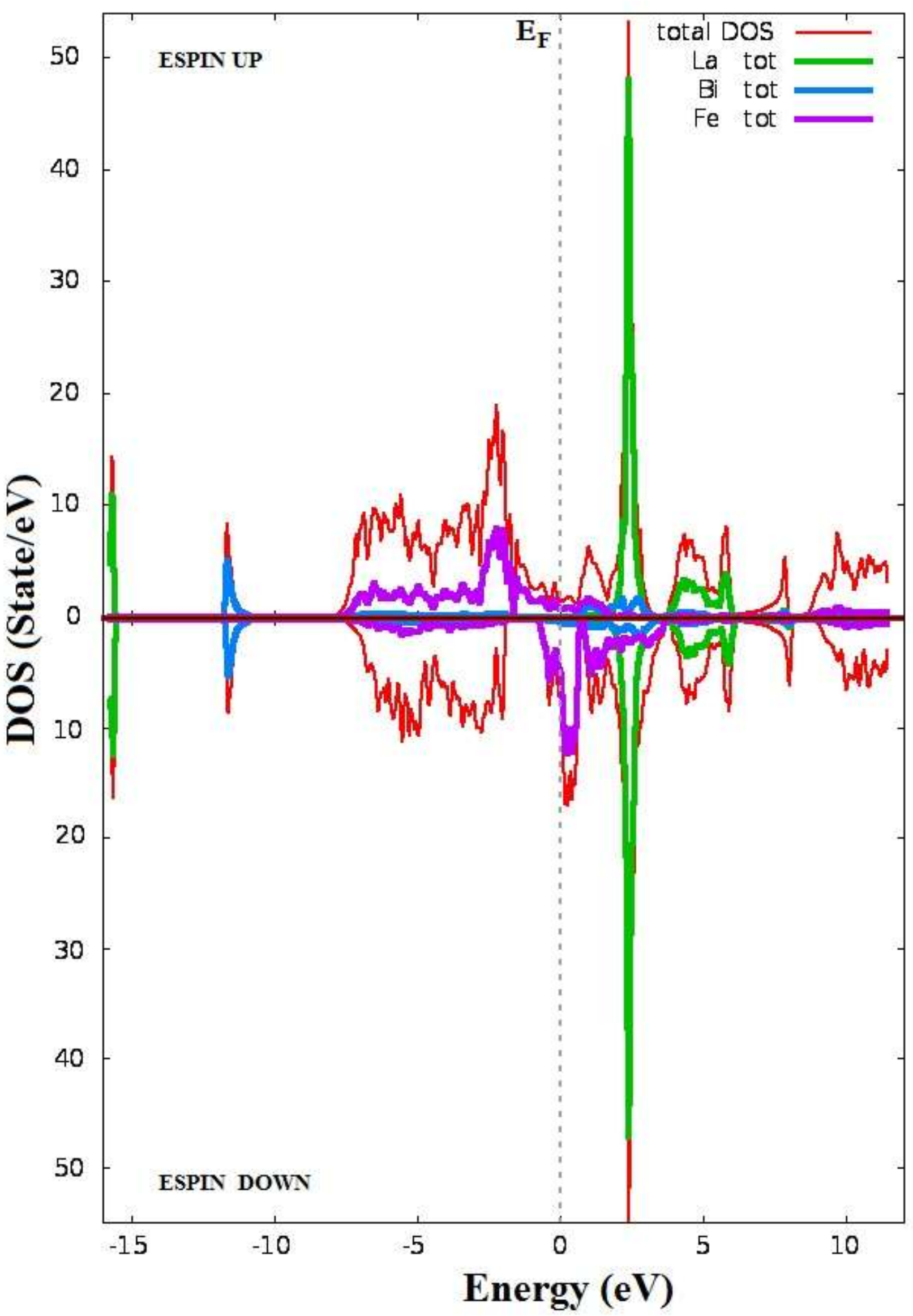

Figure 14. Total DOS calculated for the up and down spin configurations. In the picture, zero energy represents the Fermi level.

# 4 Conclusions

Samples of $LaBiFe_2O_6$ were produced by the solid state reaction recipe. Crystallographic analysis was performed by Rietveld refinement of experimental X-ray diffraction patterns, which reveal that this material crystallizes in an orthorhombic structure, belonging to the *Pnma* (#62) space group, with lattice parameters *a=5,6321 Å*, *b=7,8620* Å, *c=5,4925 Å* and cell volume of *243.81* $Å^3$. The crystal size calculated from the XRD results was of *56,4 nm*. The morphologic study through SEM images shows the occurrence of two grain sizes of *1.01* $\mu m$ and *309 nm*. The magnetization hysteresis curves suggest have a material of easy magnetization and by the classification was identified as hard magnetic, due the high values of remnant magnetization and coercive fields. The behaviour of the magnetic domains was studied from susceptibility measures as a function of temperature in the ZFC and FC procedures. For the first recipe the compound shows a metastable behaviour with second evidences a ferromagnetic feature. For a temperature of *258 K* sample exhibits an anomalous behaviour due to magnetocrystalline, shape and stress anisotropy contributions. Curves of polarization as a function of the applied voltage evidence strong dielectric loss. The dielectric constant calculated from the saturation polarization values was *81.45* for an applied frequency of *100 Hz*. This value is in agreement with that obtained from measurements of the dielectric constant which was *82,53* for a frequency of *100 Hz* and *15.66* for frequencies above *1,0 kHz*. From measurements of diffuse reflectance the energy gap was experimentally determined to be to *2,17 eV*. On the other hand, the energy gap calculated from DFT corresponds to a direct gap value of *1,3 eV*. This circumstance has to do with errors in the accurate estimation method of the energy gap, which would be necessary to introduce some corrections [48, 49]. Finally, the effective magnetic moment calculated was $13.72\ \mu_B$, which is close to the experimental result of $12,25\ \mu_B$ obtained from the Curie-Weiss analysis. The semiconductor behavior with ferromagnetic response of the $LaBiFe_2O_6$ double perovskite suggests that this material would be an excellent candidate for applications in semiconductor spin electronics due to its potential to enhance the functionality of devices that use the spin degrees of freedom of electrons, such as RAM memories and spin transistors. On the other hand, synthesis of two ferrites of La and Bi in a single crystallographic phase, maintain the ferromagnetic character observed in $LaFeO_3$ and indicate possible ferroelectric feature as a result of the presence of Bi-cations partially occupying the La-sites in the crystal structure.

**Acknowledgments.** This work was partially supported by Division of Investigations (DIB) of the National University of Colombia. One of us (J.A. Cuervo Farfán) received support by Departamento Administrativo de Ciencia y Tecnología "Francisco José de Caldas", COLCIENCIAS.